\documentclass[doublecol,linenumbers]{epl2} 

\usepackage[utf8]{inputenc}
\usepackage[T1]{fontenc}
\usepackage[english]{babel}
\usepackage{graphicx}
\usepackage{latexsym,amsmath,amssymb,amsthm}
\usepackage{makeidx}

\newcommand{\psit}{\psi_{\text{T}}}
\newcommand{\mrtwo}{$\text{M(RT)}^2$ }
\newcommand{\Eloc}{E_{\text{loc}}}
\newcommand{\psijs}{\psi_{\text{JS}}}
\newcommand{\Jee}{J_{\text{ee}}}
\newcommand{\Jep}{J_{\text{ep}}}
\newcommand{\Jse}{J_{\text{se}}}
\newcommand{\Jsp}{J_{\text{sp}}}
\newcommand{\uyuk}{u_{\text{YUK}}}
\newcommand{\Htwo}{$\text{H}_2 \,$}

\newcommand{\Ry}{\text{Ry}}
\newcommand{\psiswf}{\psi_{\text{SWF}}}
\newcommand{\psiaswf}{\psi_{\text{ASWF}}}
\newcommand{\psifswf}{\psi_{\text{FSWF}}}
\newcommand{\Jp}{J_{\text{p}}}
\newcommand{\Js}{J_{\text{s}}}

\title{On Fermionic Shadow Wave Functions for strongly-correlated multi-reference systems based on a single Slater determinant}
\shorttitle{Fermionic Shadow Wave Functions} 

\author{Francesco Calcavecchia\inst{1,2,3} \and Thomas D. K\"uhne\inst{4,5,6}}
\shortauthor{F. Calcavecchia and T. D. K\"uhne}

\institute{                    
  \inst{1} Institute of Physics, Johannes Gutenberg-University, Staudingerweg 7, D-55128 Mainz, Germany\\
  \inst{2} Graduate School of Excellence Materials Science in Mainz, Staudingerweg 9, D-55128 Mainz, Germany \\
  \inst{3} LPMMC, UMR 5493, Boîte Postale 166, 38042 Grenoble, France \\
  \inst{4} Institute of Physical Chemistry, Johannes Gutenberg-University, Staudingerweg 7, D-55128 Mainz, Germany\\
  \inst{5} Center for Computational Sciences, Johannes Gutenberg-University, Staudingerweg 9, D-55128 Mainz, Germany\\
  \inst{6} Department of Chemistry, University of Paderborn, Warburger Str. 100, D-33098 Paderborn, Germany
}
\pacs{02.70.Ss}{Quantum Monte Carlo methods}
\pacs{31.15.vn}{Electron correlation calculations for diatomic molecules}
\pacs{71.15.-m}{Methods of electronic structure calculations}

\abstract{
We demonstrate that extending the Shadow Wave Function to fermionic systems facilitates to accurately calculate strongly-correlated multi-reference systems such as the stretched \Htwo molecule. This development considerably extends the scope of electronic structure calculations and enables to efficiently recover the static correlation energy using just a single Slater determinant. 
}

\begin{document}

\maketitle

\section{Introduction}
\label{sec:introduction}

One of the most outstanding problems of computational physics and quantum chemistry is the ability to devise a quantitatively precise, yet computationally tractable, method to accurately break a chemical bond across an entire reaction coordinate. A particularly simple example is the \Htwo molecule, in particular when the covalent bond between the H atoms is stretched. Effective single-particle theories, such as the widely employed Hartree-Fock (HF) or Density Functional Theory (DFT) methods, describe the covalent bond well, but the energy is severely overestimated upon dissociation \cite{KochHolthausen}. This well-known problem is attributed to the multi-reference character of the stretched \Htwo molecule, or static electron correlation that arises in situations with degeneracy or near-degeneracy, as in transition metal chemistry and strongly-correlated systems in general \cite{Cohen08082008}. As a consequence, the stretched \Htwo molecule and similar problems are typically dealt with using multi-determinant wave functions \cite{MRwires}. However, for larger systems with many degeneracies, the number of determinants quickly becomes unfeasible \cite{FCIqmc1, FCIqmc2}. 

Variational Monte Carlo (VMC) is an accurate stochastic method that utilizes the full many-body wave function (WF) and permits to approximately solve the many-body Schr\"odinger equation \cite{McMillan:1965uq}. In contrast to quantum-chemical methods, where the computational complexity grows rapidly with the number of considered electrons $N$ \cite{PopleQC}, the formal scaling of quantum Monte Carlo (QMC) methods is similar to those of HF and DFT \cite{LuchowQMCreview,NeedsQMCreview,MitasQMCreview,LesterQMCreview,CeperleyQMCreview}. However, since it typically relies on HF or DFT orbitals to construct the Slater determinant (SD), it only allows to extract the vast majority of dynamic electron correlation, but suffers from exactly the same static correlation error that is characteristic for single-reference electronic structure methods \cite{MRCC}.

In the present work we demonstrate that extending the Shadow Wave Function (SWF), which was first introduced by Kalos and coworkers~\cite{PhysRevLett.60.1970, PhysRevB.38.4516}, to fermionic systems allows bypass the static correlation problem and permits to study strongly-correlated multi-reference systems within a much more efficient single-determinant scheme \cite{KalosReatto1995, PhysRevB.53.15129, PederivaFSWF, calcavecchia:junq_paper, sign_problem}. 

\section{Method} 
\label{sec:method}
Let us begin by very briefly reviewing the basic principles underlying the VMC method that 
relies on the Rayleigh-Ritz variational principle and importance sampled Monte Carlo (MC) techniques to efficiently evaluate high-dimensional integrals \cite{BinderMC, Kalos:MC, Binder:MC}. Due to the fact that the exact WF of the electronic ground state is unknown from the outset, it is approximated by a trial WF $\psit(R,\alpha)$, where $R \equiv \left( \mathbf{r}_1, \mathbf{r}_2, \dots \mathbf{r}_N \right)$ are the particle coordinates. 
The variational parameters $\alpha \equiv (\alpha_i)_{i=1,\dots n}$, which corresponds to the lowest variational energy
\begin{equation}
	E=\frac{\int dR \, \psit^*(R,\alpha) H \psit(R,\alpha) }{\int dR \, \psit^*(R,\alpha) \psit(R,\alpha) } \label{eq:E_for_VP}
\end{equation}
represents the best possible approximation of the electronic ground state within the given trial WF $\psit(R,\alpha)$. Thus, the accuracy of a VMC simulation critically depends on how well the particular trial WF mimics the exact ground state WF. 

In order to efficiently evaluate the high-dimensional integral of Eq.~\eqref{eq:E_for_VP}, it is conveniently rewritten as
\begin{equation}
	E=\frac{\int dR \, |\psit(R,\alpha)|^2 \frac{H \psit(R,\alpha)}{\psit(R,\alpha)} }{\int dR \, |\psit(R,\alpha)|^2  },
\end{equation}
which enables to compute it by means of MC methods. By sampling $M$ points from the probability density function
\begin{equation}
	\rho(R)= \frac{|\psit(R,\alpha)|^2}{\int dR \, |\psit(R,\alpha)|^2}
\end{equation}
using the \mrtwo algorithm (also known as the Metropolis algorithm) \cite{MRT2}, the variational energy can be estimated as
\begin{equation}
	E \simeq \frac{1}{M} \sum_{i=1}^M \Eloc(R_i),
\end{equation}
where
\begin{equation}
	\Eloc(R) \equiv \frac{H \psit(R,\alpha)}{\psit(R,\alpha)}
\end{equation}
is the so-called local energy. 

\subsection{Stochastic Reconfiguration}
Having shown how the high-dimensional integral of Eq.~\eqref{eq:E_for_VP} can be efficiently computed using the \mrtwo algorithm, it is still necessary determine the optimal variational parameters $\alpha$, which minimizes the variational energy. Here this is solved by devising a modified version of the Stochastic Reconfiguration (SR) method from Sorella \cite{PhysRevB.71.241103}. 
The SR scheme prescribes that the variational parameters are varied according to
\begin{equation}
	\delta\alpha_l = \lambda \sum_{k=1}^{n} f_k \left( s^{-1} \right)_{kl}, \label{eq:SR-direction}
\end{equation}
where 
\begin{equation}
	\left\{
	\begin{array}{lcl}
	s_{lk} & = & \langle O_k O_l \rangle - \langle O_l \rangle \langle O_k \rangle \\
	f_k & = & \langle H \rangle \langle O_k \rangle - \langle O_k H \rangle \\
	O_k & = & \frac{\partial}{\partial \alpha_k} \ln(\psit)
	\end{array}
	\right\}
\end{equation}
and $\langle \cdot \rangle \equiv \langle \psit | \cdot | \psit \rangle$. Once the direction in the variational parameters space that minimizes the variational energy has been computed, the step length $\lambda$ along this direction needs to be determined. Due to the fact that determining the new direction $\delta \alpha$ is computational approximately equally expensive than calculating the the variational energy, we found it convenient to start with a rather small value for $\lambda$ and continuously adjusting it on the fly during the optimization. Specifically, $\lambda$ is decreased whenever the search direction changes, which means that we are "moving too fast", and increased otherwise. 
The modified SR algorithm then reads as follows: 
\begin{enumerate}
	\item $i=0$
	\item Sample from the trial WF, which corresponds to the initial variational parameters $\alpha^{(0)}$ and estimate the $n$-dimensional vector $\delta \alpha^{(0)}$ by means of Eq.~\ref{eq:SR-direction}. Then, normalize $\delta \alpha^{(0)}$:
	$$\delta \alpha^{(0)}:=\frac{\delta \alpha^{(0)}}{|\delta \alpha^{(0)}|}$$
	\item $\lambda=\sqrt{\frac{|\alpha^{(0)}|^2}{n}}$
	\item $\alpha^{(1)}=\lambda \, \delta \alpha^{(0)}$
	\item $i:=i+1$ \label{pt:quinto_SR1}
	\item Estimate $\delta \alpha^{(i)}$ by sampling from the trial WF with variational parameters $\alpha^{(i)}$ and normalize it:
	$$\delta \alpha^{(i)}:=\frac{\delta \alpha^{(i)}}{|\delta \alpha^{(i)}|}$$
	\item If $i>2$ then $\lambda:=\lambda \left( 1 + 0.1 \times \delta \alpha^{(i)} \cdot \delta \alpha^{(i-1)}  \right)$
	\item If $i>3$ then
	\begin{equation*}
		\lambda:=\lambda \left( 0.85 + 0.3 \times \frac{|\alpha^{(i)}-\alpha^{(i-2)}|}{\sum_{j=1}^{2} |\alpha^{(i+1-j)}-\alpha^{(i-j)}| }  \right)
	\end{equation*}
	\item If $i>5$ then
	\begin{equation*}
		\lambda:=\lambda \left( 0.75 + 0.5 \times \frac{|\alpha^{(i)}-\alpha^{(i-4)}|}{\sum_{j=1}^{4} |\alpha^{(i+1-j)}-\alpha^{(i-j)}| }  \right)
	\end{equation*}
	\item $\alpha^{(i+1)}=\lambda \, \delta \alpha^{(i)}$
	\item Repeat steps 5-10 until convergence.
\end{enumerate}


The most commonly employed trial WF to describe the electronic structure within VMC is the so-called Jastrow-Slater (JS) WF and consists of a single SD that is multiplied by a simple correlation factor of the Jastrow form to recover most of the dynamic correlation effects \cite{PhysRev.98.1479}: 
\begin{equation}
	\psijs (R) \equiv \det(\phi_{\alpha}(\mathbf{r}_{\beta}^{\uparrow})) \det(\phi_{\alpha}(\mathbf{r}_{\beta}^{\downarrow})) \, \Jee(R) \, \Jep(R,Q),
\end{equation}
where $\alpha$ and $\beta$ are the row and column indexes of the SDs for the spin-up and spin-down electrons, while $\phi_{\alpha}$ are the single-particle orbitals that are typically determined by mean-field theories, such as HF or DFT. The Jastrow correlation factor \mbox{$J=e^{-\sum_{i,j} u(r_{ij})}$}, where $u(r_{ij})$ is a two-body pseudopotential, for the electron-electron and electron-proton interactions are denoted as $\Jee$ and $\Jep$, respectively. Here we have implemented the Yukawa-Jastrow pseudopotential for both $\Jee$ and $\Jep$, which is defined as
\begin{equation}
	\uyuk(r) \equiv A \frac{1 - e^{-Fr}}{r} 
\end{equation}
and where $A$ and $F$ are both variational parameters. The Yukawa-Jastrow pseudopotential is able to satisfy Kato's cusp condition, since
\begin{equation}
	\uyuk(r) \xrightarrow{\scriptscriptstyle r \to 0} AF - \frac{A F^2}{2}r + \mathcal{O}(r^2).
\end{equation}
However, we have not exploited the cusp condition to fix one of the two parameters, but instead we have determined both of them by means of the modified SR algorithm.



\subsection{Shadow Wave Function}
Any arbitrary trial WF can be systematically improved using the SWF formalism of Kalos and coworkers \cite{PhysRevLett.60.1970},  
which can be derived by applying the imaginary-time propagator $e^{- \tau H}$ that projects $\psit \not\perp \psi_{\text{GS}}$ onto the ground state WF $\psi_{\text{GS}}$. In order to demonstrate this, let us decompose the trial WF into
\begin{equation}
	\psit = \sum_{n=0}^{+ \infty} c_n \phi_n,
\end{equation}
where $\phi_n$ are the eigenfunctions of the Schr\"odinger equation, i.e. $H\phi_n=E_n\phi_n$ for all $n~\in~\mathbb{N}$, with $E_n$ indicating the associated energy eigenenvalues. Applying the imaginary-time propagator onto $\psi$, we obtain
\begin{equation}
	e^{- \tau H} \psit = \sum_{n=0}^{+ \infty} c_n e^{- \tau E_n} \phi_n.
\end{equation}
The projector $e^{-\tau H}$ causes that all excited components are exponentially decaying~\footnote{If some energy eigenvalues $E_n$ are negative, the corresponding term is exponentially increasing instead of decaying. Nevertheless, it is always possible to add an appropriately chosen constant energy-shift to $H$, such that all excited components are again exponentially decaying.}, so that eventually the ground state energy $E_0$ is projected out, i.e.
\begin{equation}
	\lim_{\tau \rightarrow \infty} e^{- \tau H} \psit = \lim_{\tau \rightarrow \infty} \sum_{n=0}^{+ \infty} c_n e^{- \tau E_n} \phi_n \propto \phi_0 .
\end{equation}
As a consequence, any arbitrary trial WF $\psit$ can be systematically improved by 
\begin{subequations}
\begin{eqnarray}
	e^{-\tau H} \psit(R) & = & \langle R | e^{-\tau H} | \psit \rangle \\
	                        & = & \int dS \, \langle R | e^{-\tau H} | S \rangle \langle S | \psit \rangle,
\end{eqnarray}
\end{subequations}
where we have introduced an integral over a complete set of Dirac deltas $| S \rangle$ and omitted the inessential normalization factor. 
Assuming that $\tau \ll 1$, we now use the Trotter formula to approximate
\begin{equation}
	e^{- \tau (K + V)} \sim e^{-\frac{\tau}{2} V} e^{-\tau K}  e^{-\frac{\tau}{2} V},
\end{equation}
where $K$ represents the kinetic term of the Hamiltonian \cite{Trotter}. Using the equality 
\begin{equation}
	\langle x | e^{- \tau K} | y \rangle = \frac{e^{-\frac{(x-y)^2}{4 \tau}}}{a},
\end{equation}
where $a$ is a normalization factor, the eventual expression for the improved trial WF reads as
\begin{equation}
	e^{-\tau H} \psit(R) = e^{- \frac{\tau}{2} V(R)} \int dS \, e^{- \frac{\tau}{2} V(S)} e^{- \frac{(R-S)^2}{4 \tau} } \langle S | \psit \rangle. \label{eq:first_way_to_derive_swf_final_eq}
\end{equation}

However, throughout our derivation we have assumed that $\tau \ll 1$. Hence, the imaginary-time propagation is rather short and the trial WF only slightly improved. For the purpose to elongate the propagation in imaginary-time and to solve the Schr\"odinger equation exactly, the described procedure needs to be applied repeatedly, which eventually results in a formalism similar to the path-integral approach \cite{Feynman:PI, Kleinert:PI}. 
Yet, there is no explicit importance sampling in path-integral MC methods \cite{CeperleyAlderQMC}. 
Therefore, following our original intention to find an improved and computational efficient trial WF, we rather truncate the projection after one step and refine the obtained functional form at the variational level.
In other words, instead of approaching the limit $\tau \rightarrow 0$, we instead substitute $\tau$ by a variational parameter $C$ in the gaussian term.
Moreover, we interpret the exponential $e^{-V(R)}$ as the Jastrow correlation factor $\Jp(R)$ for the protons and likewise $e^{-V(S)}$ as the corresponding two-body correlation term $\Js(S)$ for the shadows. The identity \mbox{$\langle S | \psit \rangle = \psit(S)$} implies that the original trial WF has to be evaluated on the shadow coordinates $S\equiv \left( \mathbf{s}_1, \mathbf{s}_2, \dots \mathbf{r}_N \right)$. The latter is particularly important for the term that dictates the symmetry of the SWF, which is a product of orbitals for a bosonic and a SD for a fermionic system, respectively. 
From this it follows that any trial WF $\psit$ can be systematically improved by shadow formalism. The resulting SWF then reads as 
\begin{equation}
	\psiswf(R) = \Jp(R) \, \int dS \, e^{-C \sum_{i=1}^{N} \left( \mathbf{r}_i - \mathbf{s}_i \right)^2 } \,  \Js(S) \, \psit(S).  \label{eq:SWF_first_way}
\end{equation}
From the discussion above it is apparent that the SWF can also be thought of as an one-step Variational Path Integral \cite{RevModPhys.67.279}.

Although the implementation of the SWF for bosons is relatively straightforward, the extension to fermionic systems is nontrivial due to the antisymmetry requirement of the WF to obey the Pauli exclusion principle. 
The natural way to devise an antisymmetric version of the SWF is to introduce a SD for each of the spins as a function of $S$, i.e. $\det(\phi_{\alpha}(\mathbf{s}_{\beta}^{\uparrow}))$ and $\det(\phi_{\alpha}(\mathbf{s}_{\beta}^{\downarrow}))$, respectively.
This results in the so-called Fermionic Shadow Wave Function (FSWF) 
\begin{eqnarray}
	\psifswf(R) &=& \Jee(R) \, \Jep(R,Q) \int dS \, e^{-C(R-S)^2} \, \Jse(S,R)  \nonumber \\
	            &\times& \Jsp(S,Q) \, \det(\phi_{\alpha}(\mathbf{s}_{\beta}^{\uparrow})) \det(\phi_{\alpha}(\mathbf{s}_{\beta}^{\downarrow})), 
\end{eqnarray}
where $\Jse(S,R)$ is the electron-shadow and $\Jsp(S,Q)$ the shadow-proton Jastrow correlation factor \cite{KalosReatto1995, PederivaFSWF, calcavecchia:junq_paper, sign_problem}. The FSWF, however, suffers from a sign problem \cite{calcavecchia:junq_paper, sign_problem}, which differs from the infamous fermion sign problem of projection QMC methods such as Green's function or diffusion MC \cite{Kalos:1974kx, Ceperley:1980vn}, but limits its applicability to relatively small systems. A simple \textit{ansatz} to bypass the sign problem is the Antisymmetric Shadow Wave Function (ASWF)
\begin{eqnarray}
	\psiaswf(R) &=& \Jee(R) \, \Jep(R,Q) \det(\phi_{\alpha}(\mathbf{r}_{\beta}^{\uparrow})) \det(\phi_{\alpha}(\mathbf{r}_{\beta}^{\downarrow})) \nonumber \\
	            &\times& \int dS \, e^{-C(R-S)^2} \, \Jse(S,R) \, \Jsp(S,Q), 
\end{eqnarray}
where $\det(\phi_{\alpha}(\mathbf{r}_{\beta}^{\uparrow}))$ and $\det(\phi_{\alpha}(\mathbf{r}_{\beta}^{\downarrow}))$ are SDs as a function of the electronic coordinates only \cite{PhysRevB.53.15129}. Even though the ASWF already includes many-body correlation effects of any order, the FSWF is superior since it accounts not only for symmetric, but moreover also for asymmetric, three-body and backflow correlation effects \cite{FeynmanBackflow, PhysRev.102.1189}. 

\section{Application to the \Htwo molecule}
The effectiveness of the various SWFs by means of the VMC method is demonstrated on the \Htwo molecule, whose Hamiltonian (in atomic units) reads as 
\begin{eqnarray}
	H &=& - \frac{1}{2} \sum_{i=1,2} \nabla^2_i + \frac{1}{|| \mathbf{r}_1 - \mathbf{r}_2 ||} + \frac{1}{|| \mathbf{q}_1 - \mathbf{q}_2 ||} - \nonumber \\
	&& \sum_{\substack{i=1,2 \\ j=1,2}} \frac{1}{|| \mathbf{r}_i - \mathbf{q}_j ||} \, ,
\end{eqnarray}
where $\mathbf{r}$ represents the electronic coordinates and $\mathbf{q}$ the protonic coordinates. Since $H$ includes the bare Coulomb potential, spin interactions are neglected. Due to the fact that the electrons of the \Htwo molecule possess antiparallel spins, the SDs can be replaced by the orbitals themselves. To that extend we have considered two possibilities. The first one is to use simple translational invariant plane waves (pw) orbitals, by setting $\phi=1$ (since $\mathbf{k}_1$=0). In this way, only the Jastrow correlation factor accounts for all the relevant physics. Alternatively, more accurate orbitals can be computed at the DFT level. Here we have employed the PWscf program of the Quantum Espresso package together with a pw cutoff of just $2~\Ry$, the bare Coulomb pseudopotential and the PBE exchange and correlation functional \cite{QEpaper, PhysRevLett.77.3865}. 
Since for the \Htwo molecule the sign problem is irrelevant, it is possible to directly employ both, the ASWF and the more accurate FSWF. The specific trial WFs we have considered in the present work are summarized in Tab.~\ref{tab:trial_wave_function}.
\begin{table*}
  \centering
  \begin{tabular}{|lc|l|}
    \hline
    \bf{JS-pw}              && $\displaystyle \Jee(R) \, \Jep(R,Q) $ \\
    \hline
	\bf{JS-DFT}             && $\displaystyle \Jee(R) \, \Jep(R,Q) \, \phi^{\text{DFT}}(\mathbf{r}^{\uparrow}) \phi^{\text{DFT}}(\mathbf{r}^{\downarrow}) $ \\
	\hline
	\bf{ASWF-pw/FSWF-pw}  && $\displaystyle \Jee(R) \, \Jep(R,Q) \, \int dS \, \Jse(S,R) \, \Jsp(S,Q)$  \\
	\hline
	\bf{ASWF-DFT}  && $\displaystyle \Jee(R) \, \Jep(R,Q) \, \phi^{\text{DFT}}(\mathbf{r}^{\uparrow}) \phi^{\text{DFT}}(\mathbf{r}^{\downarrow}) \, \int dS \, \Jse(S,R) \, \Jsp(S,Q)$  \\
	\hline
	\bf{FSWF-DFT}  && $\displaystyle \Jee(R) \, \Jep(R,Q) \, \int dS \, \Jse(S,R) \, \Jsp(S,Q) \, \phi^{\text{DFT}}(\mathbf{s}^{\uparrow}) \phi^{\text{DFT}}(\mathbf{s}^{\downarrow}))$  \\
	\hline
  \end{tabular}
  \caption{Employed trial WFs for the unpolarized \Htwo molecule.}
  \label{tab:trial_wave_function}
\end{table*}


\section{Results and Discussion} 
\label{sec:results}
\begin{figure}
	\centering
		\includegraphics[width=8.5cm]{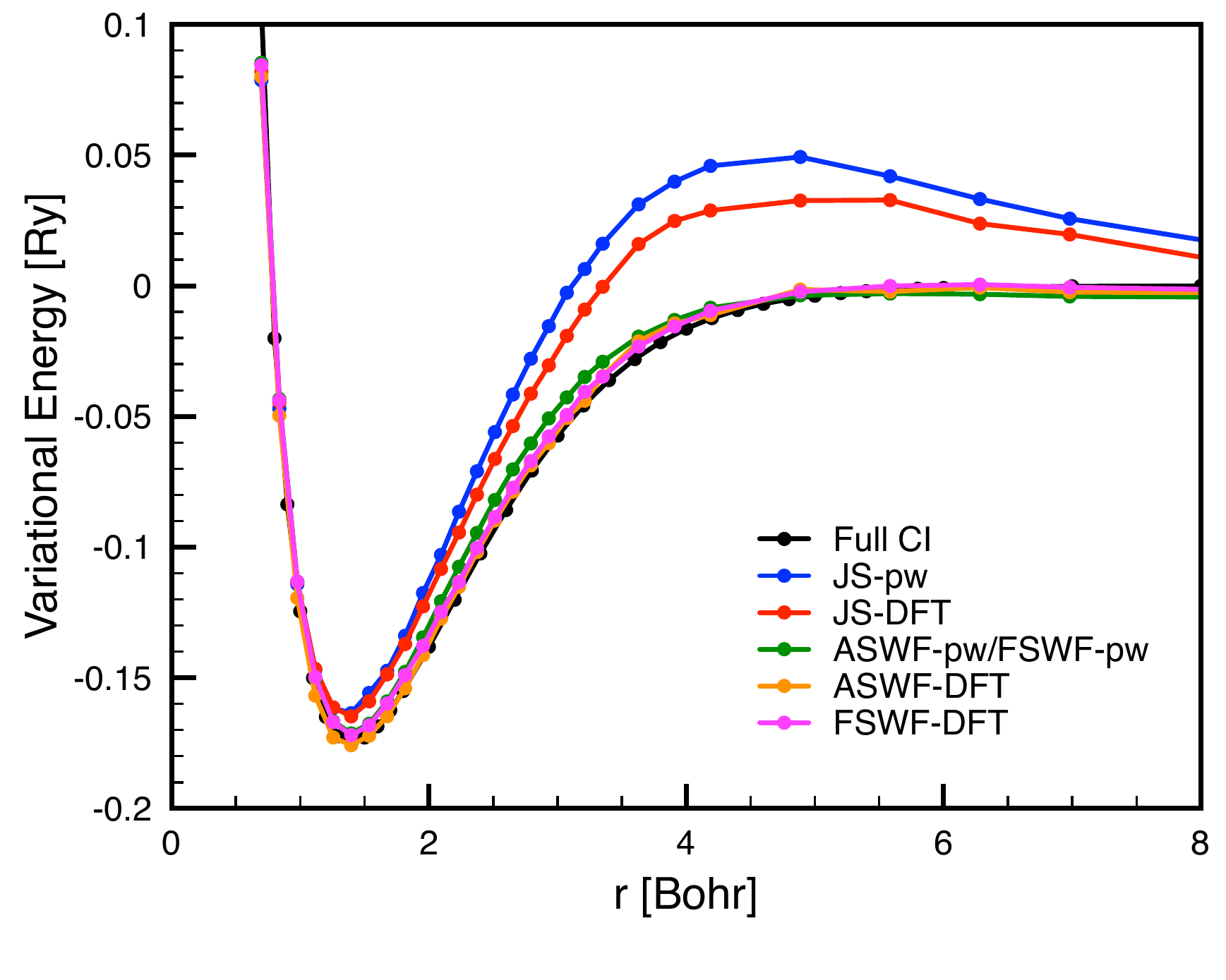} 
	\caption{Variational binding energy curves of the \Htwo molecule, obtained by subtracting the variational energy with $r=10~\text{Bohr}$ (dissociated energy) to the total variational energy. We remark that, as a consequence of such offset, a lower binding energy does not imply a lower variational energy, since such offset is different for each trial wave function.}
	\label{fig:Bond_energy}
\end{figure}

\begin{figure}
	\centering
		\includegraphics[width=8.5cm]{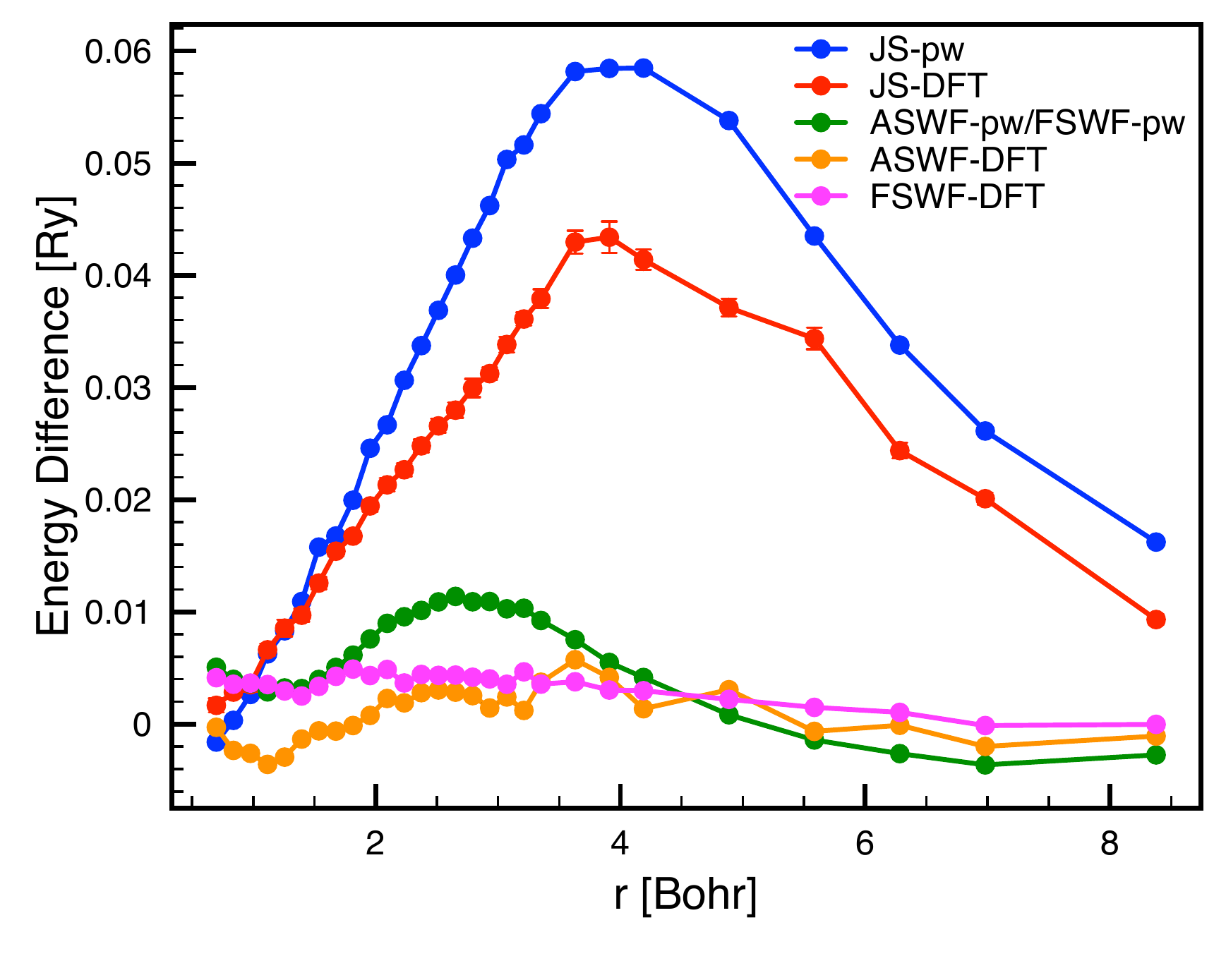}
	\caption{Difference between the binding energy curves of the \Htwo molecules as obtained using the various trial WFs and the exact full CI reference \cite{Kolos1:h2_dissociation_energy,Kolos2:h2_dissociation_energy}.}
	\label{fig:Difference-bond_energy}
\end{figure}

In Fig.~\ref{fig:Bond_energy} the \Htwo binding energy curves as obtained using the various trial WFs are shown together with the exact full configuration interaction (CI) reference data for comparison \cite{Kolos1:h2_dissociation_energy,Kolos2:h2_dissociation_energy}.
The differences between the considered trial WFs and the exact full CI reference is plotted in Fig.~\ref{fig:Difference-bond_energy}. 
It is evident that although the DFT orbitals are generally superior to simple pw orbitals, in either case the commonly employed JS trial WF substantially overestimates the potential energy of the \Htwo molecule upon dissociation. 
In contrast, the ASWF and FSWF are able to quantitatively reproduce the exact full CI reference data. As can be seen in Fig.~\ref{fig:Difference-bond_energy}, in particular for the FSWF, the energy difference is not only very small, but more importantly approximately constant.  
This is to say that despite the strong multi-reference character of the stretched \Htwo molecule, the FSWF is not only capable to recover the dynamic but also the static correlation energy, using a single SD only. 
As such, even with a nearly minimal basis set, the FSWF is very competitive with exact or highly accurate, but computational much more demanding, electronic structure techniques such as full CI QMC \cite{FCIqmc1, FCIqmc2}, projection QMC \cite{Kalos:1974kx, Ceperley:1980vn} and the Coupled Cluster method, which for the \Htwo molecule is exact and equivalent to the full CI method \cite{RevModPhys.79.291}.

\section{Conclusions} 
\label{sec:conclusions}
To summarize, in the present work we have extended the SWF to fermionic systems and derived the underlying connection between the SWF formalism and projection QMC methods via the imaginary-time propagator. 
Moreover, we have demonstrated that the commonly used JS-DFT trial WF is able to accurately describe the covalent \Htwo bond, but fails to recover most of the static correlation energy of the stretched dimer, which possess a sizable multi-reference character and typically would require the usage of a multi-determinant WF. However, the ASWF and especially the FSWF permits to study strongly-correlated multi-reference systems and is able to quantitatively reproduce the exact \Htwo binding energy curve within an very efficient single-determinant QMC method. 



\acknowledgments
T.D.K acknowledges financial support from the IDEE project of the Carl Zeiss Foundation, as well as the Gauss Center for Supercomputing (GCS) for providing computing time through the John von Neumann Institute for Computing (NIC) on the GCS share of the supercomputer JUQUEEN at the J\"ulich Supercomputing Centre (JSC).
F.C. would like to acknowledge Markus Holzmann for useful comments and the Nanosciences Foundation of Grenoble for financial support.

\addcontentsline{toc}{chapter}{\bibname}

\end{document}